# About the inconsistency between Bohr-Wheeler's transition-state method and Kramers' escape rate in nuclear fission

Karl-Heinz Schmidt[*]

*GANIL, BP 55027, 14075 Caen cedex 5, France*

**Abstract:** The problem of an apparent inconsistency between the fission rates derived on the basis of Bohr-Wheeler's transition-state method and Kramers' dynamical model of nuclear fission, first pointed out by Strutinsky in 1973, is revisited. The study is based on studying the features of individual trajectories on the fission path.

**Introduction**

Fission is one of the most important decay channels of heavy nuclei, both cold and excited. It finally limits the upper end of the chart of the nuclides. Besides the direct scientific interest on the stability of heavy nuclei, fission is assumed to play an important role in the nucleosynthesis by the astrophysical r-process. Moreover, the energy released in the fission process feeds the energy production in nuclear power plants. It is essential in all these fields to estimate the fission-decay rate in the most realistic way.

The theoretical methods used nowadays for this purpose date back to two centennial theoretical works, which appeared only a few months after the discovery of nuclear fission. Bohr and Wheeler [1] based their approach on the statistical model of nuclear reactions, however, with an important variant: Instead of considering the number of final states in the separate fission fragments, they introduced the concept of transition states in the configuration with the lowest number of states on the fission path. Bohr and Wheeler assumed that the transition states at the barrier are populated in thermal equilibrium. Another approach, which considered the reduction of the fission rate due to friction effects, mostly by insufficient delivery of trajectories close to the barrier position or with energies close to the fission barrier for low and high viscosity, respectively, was presented by Kramers [2]. Both models do not consider transient effects. These will not be the subject of this contribution neither.

A systematic discrepancy between transition-state values of the fission-decay width derived on the basis of the Bohr-Wheeler model on the one hand and on the basis of Kramers' model on the other hand has been pointed out by Strutinsky in 1973 [3]. The present contribution intends to re-visit this problem. The motivation comes from the fact that Kramers' dynamic approach does not explicitly investigate the evolution of individual systems as a function of time, since he based his model on the integral transport equation of the Fokker-Planck type [a]. Indeed, great progress in the understanding

---

[*] Previous address: GSI, Planckstr. 1, 64291 Darmstadt, Germany
[a] Detailed considerations on Kramers' model can be found in P. Hänggi et al., Rev. Mod. Phys. 62 (1990) 251,

of dynamical aspects of fission has been achieved after explicitly considering the trajectories of individual systems on the nuclear potential-energy surface by studying the differential form of the equation of motion, namely the Langevin equations [4]. Due to the presence of the random force, it is not expected that the exact results of the Langevin equations can be formulated in terms of an analytical expression. Instead, it is our aim to draw some conclusions from revealing general principles and asymptotic tendencies.

**Basic ideas**

This chapter gives a short review on the three theoretical publications, which basically determine the present understanding on estimating the nuclear fission-decay width.

*Bohr and Wheeler*

The theoretical estimate of the transition-state fission decay width $\Gamma_f$ according to Bohr and Wheeler [1] can be formulated as

$$\Gamma_f = \frac{1}{2\pi} \frac{1}{\rho_{CN}(E)} \int_0^{E-B_f} \rho_{sad}(E - B_F - \varepsilon)\,d\varepsilon \quad , \tag{1}$$

where $\rho_{CN}(E)$ is the level density of the compound nucleus at excitation energy $E$. In the spirit of this approach, $\rho_{CN}(E)$ takes into account all states of the compound nucleus. $\rho_{sad}$ is the level density at the saddle point with a constraint on the fission distortion. These states, which are in principle included in the total number of states of the compound nucleus $\rho_{CN}$, are assumed to be unstable. The relative weight of these transition states determines the probability that the system ends up in fission. $B_f$ is the height of the fission barrier, and $\varepsilon$ represents the kinetic energy associated with the fission distortion. (For simplicity, the influence of angular momentum is not considered here.) Figure 1 depicts the basic idea of the transition-state model. Equation (1) allows to consider the properties of the level density of the compound nucleus and at the barrier in full detail. Therefore, features of nuclear structure like shell effects, pairing correlations and collective modes can be included properly.

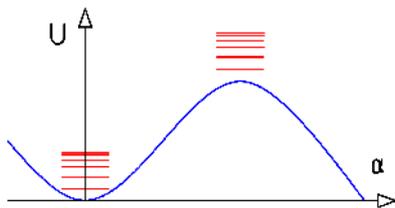

Fig. 1: Schematic representation of the basic idea of the transition-state model of Bohr and Wheeler. The potential $U$ on the fission path is drawn as a function of a suitable deformation parameter $\alpha$. The states of the compound system in the quasi-bound region around the ground state and the states at the fission barrier are indicated.

A closed expression has been derived by approximating the nuclear level density with a unique

---

however, not in direct relation to the transition-state model of Bohr and Wheeler.

constant-temperature formula

$$\rho \propto \exp(E_{intr}/T) \qquad (2)$$

as a function of the intrinsic energy $E_{intr}$ above the ground state and the fission barrier, respectively. The temperature $T$ does not depend neither on excitation energy nor on deformation: The fission-decay width of the transition-state model reduces to the following expression [5,6]:

$$\Gamma_f = \frac{T}{2\pi} e^{-B_f/T} \quad . \qquad (3)$$

The term $T$ in the pre-exponential factor results from the integration of the level density at saddle. Since the contribution to the integral is largest at the highest excitation energy, $T$ represents here the inverse logarithmic slope of the level density close to the energy $E-B_f$ above the saddle. The term $T$ in the exponent denotes the mean inverse logarithmic slope of the level density in the range between $E-B_f$ and $E$ above the ground state.

*Kramers*

Kramers' dynamical model is based on the description of the fissioning nucleus by the Fokker-Planck type equation of motion [7] on the basis of canonical thermodynamics. The model considers the collective motion along the one-dimensional fission path coupled to a heat bath with fixed temperature. The population of the excited states of the system in the fission degree of freedom is governed by Boltzmann statistics. The potential on the fission path resembles an harmonic oscillator around the ground state and an inverted parabola around the fission barrier. Both parabolas are joined smoothly. The motion in fission direction is described by a damped harmonic oscillator around the ground state and an inverted damped harmonic oscillator around the fission barrier.

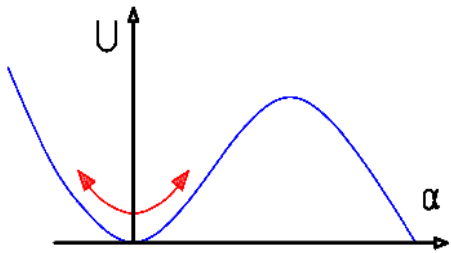

Fig. 2: Schematic representation of Kramers' dynamical model. The potential $U$ on the fission path is drawn as a function of a suitable deformation parameter $\alpha$. The oscillation around the ground state is indicated.

Kramers' result for the transition-state value of the fission-decay width is [a]:

$$\Gamma_K^{stat} = \frac{\hbar \omega_{gs}}{2\pi} e^{-B_f/T} \qquad (4)$$

Including dynamical effects, Kramers obtained the following expression for the fission-decay width

---
[a] In his original paper, Kramers uses the symbol $\omega$ to denote the frequency $f = \omega/(2\pi)$.

in the over-damped case:

$$\Gamma_K = \hbar \frac{\omega_{gs} \omega_{sad}}{2\pi\beta} e^{-B_f/T} = \Gamma_K^{stat} \frac{\omega_{sad}}{\beta} \quad . \tag{5}$$

$\beta$ is the nuclear friction coefficient, and $\omega_{sad}$ is the frequency of the harmonic oscillator corresponding to the inverted potential at the barrier[a]. The solution for the under-damped case is:

$$\Gamma_K = \hbar \beta \frac{B_f}{T} e^{-B_f/T} = \Gamma_K^{stat} \frac{\beta}{\omega_{gs}} \frac{B_f}{T} \tag{6}$$

The factors $\omega_{sad}/\beta$ and $\beta/\omega_{gs} \cdot B_f/T$ for the over-damped and the under-damped case, respectively, in equations (5) and (6) quantify the reduction of the fission-decay width due to dynamical effects. In the case of large $\beta$, the fission-decay width decreases like $1/\beta$ with increasing $\beta$ due to insufficient delivery of trajectories close to the barrier position. In the opposite case, the fission-decay width decreases proportional to $\beta$ with decreasing $\beta$ due to insufficient delivery of trajectories with energies close to the fission-barrier height. The possibility that the system returns into the quasi-bound region after having passed the barrier is also taken into account.

Due to the lack of suited mathematical tools for solving the Fokker-Planck-type equations for complex cases, it is difficult to extend Kramers' model, e.g. to consider structural effects of the nuclear level density. Therefore, Kramers' transition-state value, equation (4), is often replaced by the result of the Bohr-Wheelers approach, expressed by equation (1).

*Strutinsky*
Obviously, the transition-state values of the fission-decay width derived from the Bohr-Wheeler model in the closed-form approximation of equation (3) on the one hand and Kramers' result on the other hand (equation 4) differ by a factor $T/(\hbar\omega_{gs})$. Strutinsky came to the conclusion that the discrepancy is explained as due to the fact that the difference in the number of stationary collective states in the initial and in the transitional states was erroneously ignored by Bohr and Wheeler. Strutinsky suggested that the fission-decay width derived with the Bohr-Wheeler transition-state method should generally be multiplied with the factor $\hbar\omega_{gs}/T$ [3]. This way, the number of states $N_0 = \rho_0(E) dE$ at energy $E$ and fixed shape is replaced by the value $N_1 = \frac{dE}{2\pi\hbar} \int dq \int dp_q \, \rho(E - E_{coll})$, which is corrected for the energy $E_{coll}$ stored in the collective mode.

---

a   Kramers also derived a more general expression for the over-damped case, which is valid down to critical damping:

$$\Gamma_K = \Gamma_K^{stat} \cdot \left( \sqrt{1 + \left(\frac{\beta}{2\omega_{sad}}\right)^2} - \frac{\beta}{2\omega_{sad}} \right)$$

This argumentation seems to us slightly misleading in two aspects. First, using the term stationary collective states suggests that the inconsistency noticed by Strutinsky is restricted to the under-damped case, where stationary collective states in form of harmonic oscillations are present. This is in conflict with Strutinsky's suggestion to generally apply the correction factor $\hbar\omega_{gs}/T$, for any magnitude of the nuclear viscosity. Secondly, we think that Bohr and Wheeler's transition-state rate is correct, if realistic level densities for the compound nucleus and for the saddle-point configuration are used, as will be discussed in detail below. The usual way for evaluating this quantity is based on the independent-particle model of the nucleus: The states of the compound nucleus are derived from considering all possible combinations of single-particle excitations in the given excitation-energy range [8]. Since the excitation energy, the angular momentum and the parity of an isolated nucleus are preserved, the number of states is considered under the condition of fixed values for these three quantities. When regarding the motion of all individual nucleons, all degrees of freedom of the nucleus are exhausted, and, therefore, the independent-particle model gives a realistic estimate of the total number of states. This includes also collective states, in particular vibrational states, which include oscillations in fission direction. However, residual interactions act on the energies of some states, and thus the number of states at a given excitation energy, in particular at low energies, is better reproduced, if residual interactions are taken into account, see e.g. [9,10,11]. The number of states $\rho_{sad}$ in the saddle-point shape is understood having a constraint on the motion associated with the fission distortion. In Kramers' approach, this is the elongation degree of freedom.

**Characteristics of individual trajectories**

The equation of motion of the Langevin type allows calculating individual trajectories on the potential-energy surface numerically. In the case of the scenario behind Kramers' model, the trajectory is the motion on the one-dimensional fission path.

As an illustration we present two characteristic trajectories resulting from numerical calculations using the Langevin equations. One can clearly see almost undisturbed oscillations with slowly fluctuating amplitude in the under-damped case and the erratic dissipative character of the motion in the over-damped case.

We introduce a wall at the deformation of the fission barrier, which reflects the trajectories that otherwise would overcome the barrier. In this way, it is assured that the system is bound, and the net flux over the barrier is zero. Thus, the population of states along the deformation path corresponds to equilibrium and thus to the transition-state picture. Since the potential is symmetric close to the top of the barrier, this scenario is equivalent to the assumption of thermodynamical equilibrium on both sides of the barrier, but it avoids the need for defining the properties of the system far outside the quasi-bound region. Using this scenario, we study the behaviour of the system in the cases of very large and very small viscosity.

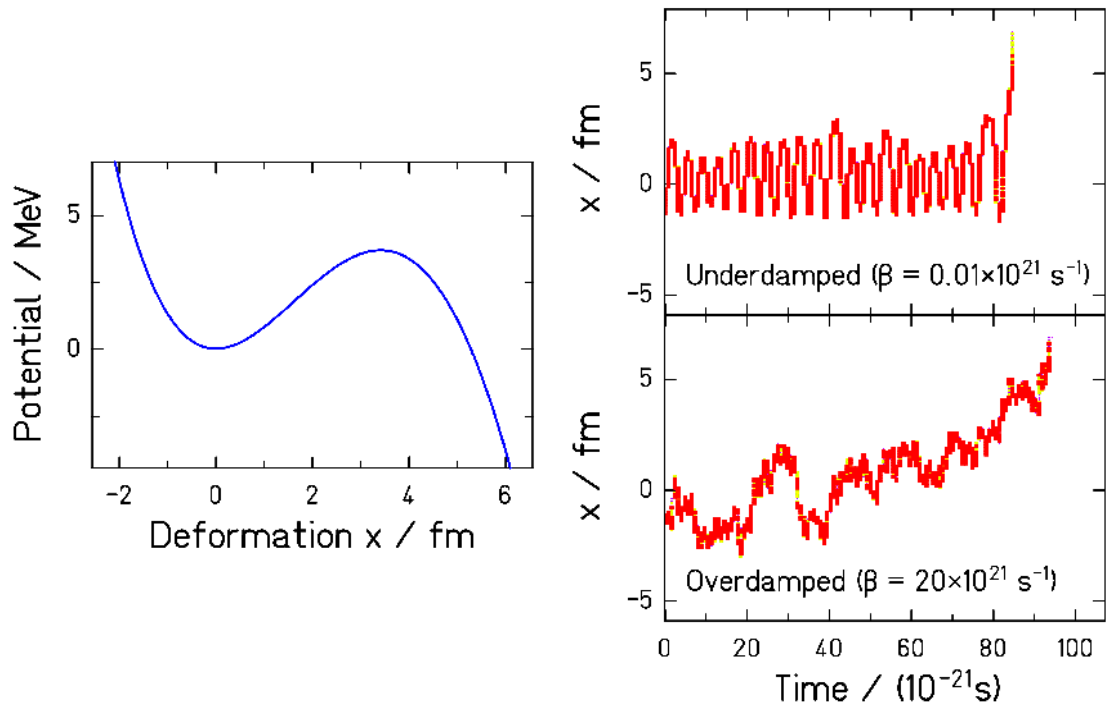

Figure 3: Results of numerical calculations using the Langevin equations. Two typical trajectories of a sample nucleus with a fission barrier of 3.7 MeV and A = 248 are shown for a strongly under-damped and a strongly over-damped case, respectively, in the right part of the figure. The potential along the fission path is sketched in the left part of the figure.

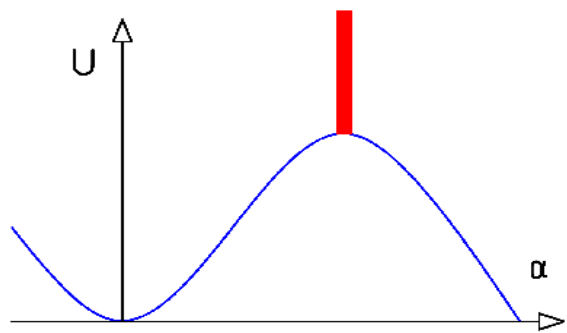

Figure 4: Basic idea how to derive the equilibrium population of transition states from the characteristics of individual trajectories: The trajectories are reflected by a wall when they reach the fission barrier. This way, the system is bound, and the net flux across the fission barrier is zero.

*Over-damped case*

In the case of large viscosity, the motion of the nucleus along the fission path is strongly over-damped, it has the characteristics of a diffusion process. It is well known as a general principle that in this case *the system spends the same time in any state accessible at the given total energy*. This is equivalent to the basic feature of the statistical model that *every state at the given total energy is populated with the same probability*. It appears that this is the basic assumption of Bohr and Wheeler for determining the population probability of the transition states. Thus, there is no indication for a discrepancy between the concepts of the two models for deriving the transition-state fission rate.

Since the diffusive motion of the system adapts fast to its local properties, details of the level density at the barrier directly and fully determine the fission flux. The flux is directly proportional to the exact number of states at saddle. Thus, equation (1) with realistic level densities gives a good estimate of the transition-state value of the fission-decay width, if $\rho_{CN}$ represents the total level density of the compound nucleus, including all its degrees of freedom and if $\rho_{sad}$ represents the density of the levels at the barrier with a constraint on the fission distortion. The motion at the barrier associated with the fission distortion is taken into account by the integral over the energy above the barrier. The effect of dissipation can additionally be considered by Kramers' dynamical model.

*Under-damped case*
In the case of zero viscosity, the motion in fission direction is a periodic oscillation with the frequency $\omega_{gs}$, very similar to an harmonic oscillator, since the shape of the potential is very similar to a parabola. If the coupling to the intrinsic nuclear degrees of freedom is weak, individual oscillations appear almost unperturbed, but the energy stored in the oscillation and thus its amplitude is slowly varying [2]. If the temperature $T$ of the heat bath is assumed to be constant, the probability for finding a specific value of the oscillator energy $E_{osc}$ is determined by the Boltzmann distribution:

$$P(E_{osc}) = T^{-1} \exp(-E_{osc}/T) \tag{7}$$

Considering that each oscillation with an energy that exceeds the fission barrier $B_f$ is reflected at the wall, and it is thus counted as a fission event in the transition-state picture, the transition-state value of the fission-decay width is given by

$$\Gamma_f = \frac{\hbar \omega_{gs}}{2\pi} \int_{B_f}^{\infty} T^{-1} \exp(-E_{osc}/T) dE_{osc} = \frac{\hbar \omega_{gs}}{2\pi} \exp(-B_f/T) \tag{8}$$

which is identical to Kramers' result.

It is the temperature, i.e. the inverse of the logarithmic slope of the level density as a function of energy, averaged over the whole oscillation, which determines the Boltzmann factor in equation (8) and thus the fission rate. In a microcanonical picture, the temperature of the 'heat bath' is also a slowly varying quantity, which is anticorrelated to the amplitude of the oscillation. This means that the exponential function in equation (8) will be replaced by a slightly different function, which might be determined numerically.

**General considerations**

The inversion of the curvature of the potential along the fission path between the ground state and the fission barrier, which is behind the phase-space factor proposed by Strutinsky, is not the only reason for a systematic difference of the level densities $\rho_{CN}$ and $\rho_{sad}$. Also the shape of the nucleus has been recognized to have sizeable influence on the level density. Different theoretical models predict an increase of the level-density parameter [12,13,14,15] e.g. with increasing surface. The level-density parameter $a_f$ at the barrier is estimated to be typically larger by a few percent compared to the level-density parameter $a_n$ for the spherical shape.

It is interesting to note that the phase-space factor has about the same effect as a reduction of the level-density parameter at the barrier. Figure 5 shows that the increase of the level density of the compound nucleus by the factor $T/(\hbar\omega_{gs})$ is quantitatively very similar to the increase of the level density at the barrier with excitation energy, corresponding to $a_f/a_n = 1.02$. In addition, one should consider that the increase of the level-density parameter with increasing deformation tends to decrease the effective value of $\hbar\omega_{gs}$ with increasing energy and angular momentum [16]. This would lead to an even stronger increase of the dashed line in figure 5 with increasing energy. This means that the increase of the fission probability due to a deformation-dependent level-density parameter is appreciably reduced or even compensated by the decrease of the fission probability due to the phase-space factor.

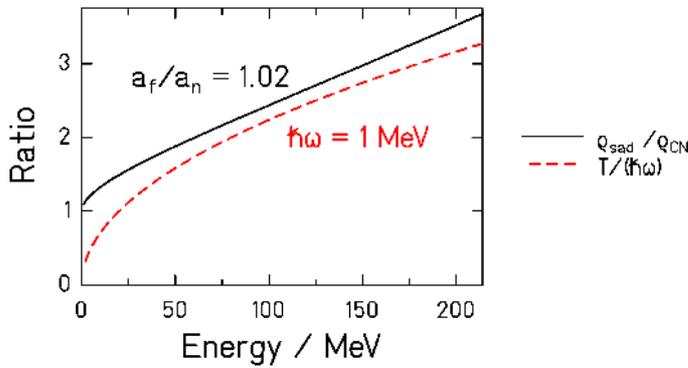

Figure 5: Full curve: Ratio $\rho_{sad}/\rho_{CN}$ of the level densities calculated as $\rho_{sad}=\exp(2\sqrt{a_f E})$ (at saddle) and $\rho_{CN}=\exp(2\sqrt{a_n E})$ (of the compound nucleus). Dashed curve: Correction factor T/($\hbar\omega$). The calculations have been performed for a nucleus with mass number $A = 220$. The other parameters are $\hbar\omega = 1$ MeV, $a_n = A/11$, $a_f = a_n * 1.02$. The temperature is given by $T=\sqrt{E/a_n}$ .

The theoretical predictions for $a_f/a_n$ have often been compared with experimental data via dedicated model calculations. In most cases, the phase-space factor was not applied, and the additional reduction of the fission-decay width due to dissipation was not considered in these investigations. Some particularly careful studies have been made by Jing et al. [17] on the basis of fission probabilities of neighbouring osmium isotopes in $^3$He-induced reactions and by Siwek-Wilczynska et al. [18] on the basis of the fusion-evaporation reactions $^{208}$Pb($^{16}$O,xn) and $^{236}$U($^{12}$C,xn). In these reactions, induced by light projectiles, the angular momentum and other complex entrance-channel effects should be small. The common conclusion of these studies is that there is no room for including the phase-space factor proposed by Strutinsky or the Kramers factor accounting for a dynamical hindrance of fission, if they differ appreciably from one.

On the other hand, an analysis of oxygen- and fluor-induced reactions with targets from $^{159}$Tb to $^{197}$Au with the phase-space factor, the Kramers factor and other higher-order corrections of the transition-state level density included was able to reproduce measured fission- and evaporation-residue cross sections as well as pre-scission neutron multiplicities without the need for variations of the dissipation strength with excitation energy [19].

**Summary**

We think that the result of the above discussion can be summarized as follows:
1. There is no conceptional inconsistency between the Bohr-Wheeler transition-state model and Kramers' dynamical model as far as the transition-state value of the fission-decay width is concerned (equation 1 and equation 4), if $\rho_{CN}$ is the density of all states of the compound nucleus and $\rho_{sad}$ denotes the density of transition states above the saddle in equation (1) with a constraint on the fission distortion.
2. When separating the total nuclear phase space in the one-dimensional motion in fission direction on the one hand and the other degrees of freedom on the other hand, the phase space in fission direction is confined by a parabolic potential around the ground state, while this confinement is not present at saddle. The occupied phase-space volume of the compound nucleus in fission direction in units of h is given by [3]

$$\frac{1}{h} \int dq \int dp_q = \frac{\int_0^\infty \exp(-E/T) dE}{\hbar \omega_{gs}} = \frac{T}{\hbar \omega_{gs}} \qquad (9)$$

   Due to the parabolic confinement around the ground state, there is a systematic difference in phase space between the compound nucleus and the configuration at the saddle point. This is the reason why Strutinsky proposed to apply the factor $\hbar \omega_{gs}/T$ to the fission-decay width, if the same theoretical level-density formula is used for $\rho_{CN}$ and $\rho_{sad}$, assuming that there is no sensible difference in the other degrees of freedom.
3. The discrepancy between the closed form of the transition-state decay width (equation 3) and Kramers' expression (equation 4) emerges from the use of the same level-density formula for the number of states of the compound nucleus $\rho_{CN}$ and for the number of states $\rho_{sad}$ above the saddle when deriving equation (3).
4. The phase-space volume and thus the number of quantum states does not depend on viscosity. Therefore, the application of the correction factor $\hbar \omega_{gs}/T$, if justified, is not restricted to the existence of true stationary collective oscillatory states, that means to the under-damped case.
5. When using realistic (e.g. experimental) level densities for $\rho_{CN}$ and $\rho_{sad}$ in expression (1),

the correction factor $\hbar\omega_{gs}/T$ should not be applied to the fission-decay width.

Besides the phase-space factor suggested by Strutinsky, there are several other reasons for systematic differences of the level-densities $\rho_{CN}$ and $\rho_{sad}$, which determine the transition-state fission rate. One of these is the expected increase of the level-density parameter with deformation. The enhancement of the fission probability due to this latter effect is appreciably reduced or even compensated by the phase-space factor over the whole excitation-energy range. Different attempts to pin down the systematic differences of $\rho_{CN}$ and $\rho_{sad}$ are still contradictory. This implies that there is still an appreciable uncertainty in the application of pre-exponential factors in the description of the nuclear level density.

**Acknowledgement**

I thank David Boilley for careful reading of the manuscript. Fruitful discussions with Christelle Schmitt and Alexander Karpov are gratefully acknowledged.

1. N. Bohr, J. A. Wheeler, Phys. Rev. C 56 (1939) 426
2. H. A. Kramers, Physica 7 (1940) 284
3. V. M. Strutinsky, Phys. Lett. B 47 (1973) 121
4. Y. Abe, C. Grégoire, H. Delagrange, J. Phys. (Paris), Colloq. 47 (1986) C4-329
5. Y. Fujimoto, Y. Yamaguchi, Progr. Theor. Phys. Japan 5 (1950) 76
6. R. Vandenbosch, J. R. Huizenga, Nuclear Fission, Academic Press, New York, 1973
7. S. Chandrasekhar, Rev. Mod. Phys. 15 (1943) 1
8. J. R. Huizenga, L. G. Moretto, Ann. Rev. Nucl. Sci. 22 (1972) 427
9. A. V. Ignatyuk, Yad. Fiz. 21 (1975) 20 (Sov. J. Nucl. Phys. 21 (1975) 10)
10. T. Dossing, A. S. Jensen, Nucl. Phys. A 222 (1974) 493
11. K. Langanke, Nucl. Phys. A 778 (2006) 233
12. A. V. Ignatyuk, M. G. Itkis, V. N. Okolovich, G. N. Smirenkin, A. S. Tishin, Sov. J. Nucl. Phys. 21 (1975) 612
13. W. Reisdorf, J. Toke, Z. Phys. A 302 (1981) 183
14. J. Tocke, W. J. Swiatecki, Nucl. Phys. A 372 (1981) 141
15. A. V. Karpov, P. N. Nadtochy, E. G. Ryabov, G. D. Adeev, J. Phys. G: Nucl. Part. Phys. 29 (2003) 2365
16. J. P. Lestone, Phys. Rev. C 51 (1995) 580
17. K. X. Jing, L. W. Phair, L. G. Moretto, Th. Rubehn, L. Beaulieu, T. S. Fan, G. J. Wozniak, Phys. Lett. B 518 (2001) 221
18. K. Siwek-Wilczynska, I. Skwira, J. Wilczynski, Phys. Rev. C 72 (2005) 034605
19. S. G. McCalla, J. P. Lestone, Phys. Rev Lett. 101 (2008) 032702